\documentstyle[prb,aps,multicol,epsf]{revtex}
\input psfig

\def\be{\begin{equation}}
\def\ee{\end{equation}}

\def\sqr#1#2{{\vcenter{\vbox{\hrule height .#2pt
      \hbox{\vrule width .#2pt height#1pt \kern#1pt
      \vrule width.#2pt}
      \hrule height.#2pt}}}}

\begin{document}
\bibliographystyle{simpl1}

\title{Absence of singular superconducting fluctuation corrections
to thermal conductivity}

\author{Douglas R Niven and Robert A Smith}

\address{School of Physics and Astronomy, University of
Birmingham, Edgbaston, Birmingham B15 2TT, ENGLAND}

\maketitle
\bigskip

\begin{abstract}
We evaluate the superconducting fluctuation corrections to thermal conductivity
in the normal state which diverge as $T$ approaches $T_c$. We find zero total
contribution for one, two and three-dimensional superconductors for arbitrary
impurity concentration. The method used is diagrammatic many-body theory, and
all contributions -- Aslamazov-Larkin (AL), Maki-Thompson (MT), and 
density-of-states (DOS) -- are considered. The AL contribution is convergent,
whilst the divergences of the DOS and MT diagrams exactly cancel.
\end{abstract}
\bigskip

The discovery of the high-$T_c$ superconductors has led to a renewed 
interest\cite{Lark01} in superconducting fluctuation corrections to normal 
state transport properties\cite{Skoc75}. 
Whilst much of the work has focused on the electrical resistivity, 
$\rho$, several experiments\cite{Cohn92,Hous96,Hous97,Sham00}
have reported fluctuation corrections to the thermal
conductivity, $\kappa$. Since there is some dispute between 
theorists\cite{Abra70,Varl90,Varl92,Vish01,Ussi02,Sav02} as to the 
predicted magnitude of the effect, we have performed a detailed microscopic
calculation valid for all impurity concentrations. We find no divergent 
fluctuation contribution, and conclude that the experimental features seen near
$T_c$ must have some other physical origin. 

Let us try to understand the reason for the lack of singular fluctuation 
contributions to thermal conductivity. There are several processes involved, and
we will try to develop a physical picture\cite{Varl99} for each. The Aslamazov-Larkin (AL)
process involves the transfer of heat by fluctuation Cooper pairs. The corresponding
term for the electrical conductivity has the strong divergence
\begin{equation}
\label{sigAL}
\sigma_{AL} \sim (T-T_c)^{d/2-2}.
\end{equation}
The size of the contribution to thermal conductivity
can be estimated from Eq.~(\ref{sigAL}) using the 
Wiedemann-Franz law, which has the general form
\begin{equation}
\label{WF}
\kappa T\sim \left({k_BT_0\over Q_0}\right)^2\sigma,
\end{equation}
where $k_BT_0$ is the amount of heat, and $Q_0$ the electric charge, carried by the
excitations in a given system. For fluctuation Cooper pairs, $T_0\sim T-T_c$, and
$Q_0=2e$ so that
\begin{equation}
\label{kAL}
\kappa_{AL}T\sim \left({k_B(T-T_c)\over 2e}\right)^2\sigma_{AL}
\sim (T-T_c)^{d/2},
\end{equation}
which is clearly non-singular as $T\rightarrow T_c$.  The density-of-states (DOS)
correction arises from the fact that when electrons form fluctuation Cooper pairs, they
cannot simultaneously act as normal electrons; there is a corresponding decrease in the
normal state density-of-states and hence normal state thermal conductivity
\begin{equation}
\label{kDOS}
\kappa_{DOS}\sim -{n_{cp}k_B^2T^2\tau\over m}\sim -(T-T_c)^{d/2-1},
\end{equation}
where $n_{cp}\sim (T-T_c)^{d/2-1}$ is the number density of fluctuating Cooper pairs.
This term is singular for $d\le 2$, but is exactly cancelled by Maki-Thompson (MT) 
terms.
The latter terms are due to new heat transport channels opened
up by Andreev scattering processes. An electron can Andreev-scatter into a hole, and since
electrons and holes carry the same heat current, this leads to a net increase in thermal
conductivity. The amplitude for the Andreev scattering is exactly the same as for an 
electron to scatter into a fluctuation Cooper pair, so the MT and DOS terms have the same
magnitude but opposite sign, and hence cancel. These MT processes lead to a further
suppression to electrical conductivity since holes carry opposite electric charge to electrons
i.e. the MT and DOS contributions cancel for thermal conductivity and reinforce for
electrical conductivity.

Before we proceed to the details of our calculation, we present a short
history of superconducting fluctuation corrections to thermal conductivity.
They were first predicted\cite{Abra70} in 1970 by Abrahams et al in the 
diffusive regime. 
These authors concluded that the Aslamazov-Larkin (AL) terms were convergent, 
but that  density-of-states (DOS) terms led to divergent
contributions in two and one dimensions, of the form $\ln{(T-T_c)}$ and 
$(T-T_c)^{-1/2}$ respectively. They appear to have missed the cancellation
between DOS and MT contributions. Shortly afterwards fluctuation effects
with the predicted power-law behaviour
were observed\cite{Wolf71} in one-dimensional Pb-In wires. 
After this intial
work there was apparently no theoretical or experimental activity in this area
for nearly two decades. Indeed, in Skocpol and Tinkham's 1975 
review\cite{Skoc75}, thermal conductivity
is described as one of those quantities which ``have not yet benefited from
sustained interaction between theory and experiment, perhaps because such
effects are small, and hard to interpret.'' In 1990 Varlamov and 
collaborators\cite{Varl90}
predicted AL contributions with the same strong divergence found in the electrical
conductivity, $(T-T_c)^{d/2-2}$; this erroneous result appears to be due to an
incorrect treatment of the heat-current operator.
The same authors\cite{Varl92} also discussed the relative magnitudes of DOS, 
MT and AL contributions
in layered superconductors, and argued that the DOS and MT terms dominate in 
$\kappa_c$ whilst AL terms dominate in $\kappa_{ab}$. The predicted fluctuation
effects have since been seen experimentally in an YBa$_2$Cu$_3$O$_{7-\delta}$   
single crystal\cite{Cohn92}, and Bi$_2$Sr$_2$CaCu$_2$O$_8$ and 
DyBa$_2$Cu$_3$O$_{7-\delta}$ polycrystals\cite{Hous96,Hous97}. 
Excellent quantitative agreement was found between theory and 
experiment; indeed, even the predicted two- to three-dimensional crossover is
seen at roughly the predicted temperature.
Fluctuation effects have also been seen in (Nd/Y)BCO intergrowth 
crystals\cite{Sham00},
although these have not been compared in detail with theory. However there
are problems with this apparent agreement between theory and experiment.
The AL contributions have been re-analysed in two works using 
phenomenological hydrodynamic\cite{Vish01} and Gaussian 
fluctuation\cite{Ussi02} approaches, and argued
to be convergent. Very recently Savona et al\cite{Sav02} have agreed that
there is no divergent AL correction, 
but argue that there are still divergent 
DOS and MT terms; we believe that these authors have missed the cancellation between
the DOS and MT terms.

We now proceed to the details of our microscopic calculation. 
The thermal conductivity is obtained from the imaginary time heat response
kernel, $Q_{hh}(i\Omega_n)$, by analytic continuation from positive Bose Matsubara
frequencies, $\Omega_n=2\pi Tn$,
\begin{equation}
\label{Qdef}
\kappa=\lim_{\Omega\rightarrow 0} 
{Q_{hh}(i\Omega_n\rightarrow\Omega+i0)\over i\Omega T}.
\end{equation}
The diagrammatic contributions to the heat response kernel of lowest order in
perturbation theory  are detailed in Fig. 1.
The solid lines are disordered electron Green functions 
\begin{equation}
\label{Gdef}
G(k,i\varepsilon_l)={1\over i\varepsilon_l-\xi_k+{i\over 2\tau}
\hbox{sgn}(\varepsilon_l)}
\end{equation}
where $\varepsilon_l=2\pi T(l+1/2)$ is a Fermi Matsubara frequency,
$\xi_k=k^2/2m-\mu$ is the electronic excitation spectrum, and $\tau$ is the elastic
scattering time. The black dots represent heat-current vertices, which are given by
\begin{equation}
\label{Jdef}
{\bf j}_h({\bf k},\varepsilon_l,\varepsilon_l+\Omega_n)=
{{\bf k}\over 2m}i(2\varepsilon_l+\Omega_n).
\end{equation}
The shaded regions are impurity vertex renormalization which, at zero momentum, take the
form
\begin{equation}
\label{Cdef}
C(q=0,\varepsilon_1,\varepsilon_2)=\Theta(+\varepsilon_1\varepsilon_2)
+{\Theta(-\varepsilon_1\varepsilon_2)\over (|\varepsilon_1|+|\varepsilon_2|)\tau},
\end{equation}
whilst the dashed lines are single impurity renormalizations.
 The wavy lines are superconducting fluctuation propagators,
$L(q,i\omega_m)$, which for small $q$ are given by
\begin{equation}
\label{Ldef}
L(q,i\omega_m)^{-1}=N(0)\left[\ln{\left({T\over T_c}\right)}
+\psi\left({1\over 2}+{|\omega_m|\over 4\pi T}\right)-\psi\left({1\over 2}\right)
+A(\omega_m)Dq^2\right]
\end{equation}
where $N(0)$ is the electronic density-of-states per spin at the Fermi surface, 
$\omega_m=2\pi Tm$ is a Bose Matsubara
frequency, $\psi(x)$ is the digamma function, $D=v_F^2\tau/d$ is the diffusion constant,
and $A(\omega_m)$ is given by
\begin{equation}
\label{Adef}
A(\omega_m)={1\over 4\pi T}\psi'\left({1\over 2}+{|\omega_m|\over 4\pi T}\right)
-\tau\left[\psi\left({1\over 2}+{|\omega_m|\over 4\pi T}+{1\over 4\pi T\tau}\right)
-\psi\left({1\over 2}+{|\omega_m|\over 4\pi T}\right)\right].
\end{equation}
The zero-frequency fluctuation propagator, $L(q,0)$, has a $1/q^2$ divergence as
$T$ approaches $T_c$,
\begin{equation}
\label{L0def}
L(q,0)^{-1}=N(0)\left[{T-T_c\over T_c}+A(0)Dq^2\right].
\end{equation}
It is this feature which leads to divergent contributions to various physical properties
as $T$ approaches $T_c$.

Diagrams A and B of Fig. 1, in which a fluctuation propagator affects only one electron line
yields the DOS contributions; diagram C, in which a fluctuation propagator leads to interference
between electron lines, yields the MT contribution; diagram D, which possesses two fluctuation
propagators, yields the AL contributions.
Note that since the object of this paper is merely to show that there are no divergent
contributions to $\kappa$ at $T=T_c$, we have omitted all diagrams that cannot have
such divergences. In particular we have ignored all DOS and MT diagrams that have an
impurity line or ladder between the two heat current vertices. Such diagrams 
possess an extra factor of $q^2$ which removes the low-momentum singularity of the 
fluctuation propagator, $L(q,0)$. We need also consider only the lowest power of $q$
in any diagram since this will have the most divergent behavior -- we therefore set
$q=0$ in all terms except the fluctuation propagators.
Finally since all DOS and MT diagrams have only one
superconducting fluctuation propagator we can take the static limit and consider only
terms $L(q,i\omega_m)$ with zero Cooper pair frequency, $\omega_m=0$. The AL term has
two fluctuation propagators, and here we have to be more careful and keep all 
$\omega_m$ terms as there is an anomalous region of frequencies where one propagator can
have positive frequency, and the other negative frequency.

The regular parts of the DOS and MT diagrams, which come from diagrams A and C, give the
total contribution
\begin{equation}
\label{MTDOSreg}
Q^{reg}_{hh}(i\Omega_n)=
-\pi N(0)DT^2\sum_{\varepsilon_l>0}
{(2\varepsilon_l+\Omega_n)^2\over [1+(2\varepsilon_l+\Omega_n)\tau]}
\left\{ {1\over\varepsilon_l^2}+{1\over (\varepsilon_l+\Omega_n)^2}
-{2\over\varepsilon_l(\varepsilon_l+\Omega_n)}\right\}\sum_q L(q,0)
\end{equation}
The sum of three terms in the curly brackets is easily seen to be proportional to
$\Omega_n^2$ so upon analytical continuation, division by $\Omega$, and setting $\Omega$
to zero, we get zero contribution. The two DOS and one MT term have exactly cancelled
each other. Note that the same terms in the electromagnetic response 
function reinforce rather than cancel each other because the electric current vertex has
the opposite electron-hole parity to the heat current vertex (ie. holes carry opposite
charge but the same excitation energy to electrons).

The anomalous parts of the DOS and MT diagrams give total contribution
\begin{eqnarray}
\label{MTDOSan}
&&Q^{anom}_{hh}(i\Omega_n)=
{\pi N(0)D\over (1+\Omega_n\tau)^2}T^2\sum_{0<\varepsilon_l<\Omega_n}
(2\varepsilon_l-\Omega_n)^2
\left\{ {1+2\varepsilon_l\tau\over\varepsilon_l^2}
+{1+\Omega_n\tau\over\varepsilon_l^2}
-{1\over\varepsilon_l^2}
+{1+\Omega_n\tau\over\varepsilon_l(\Omega_n-\varepsilon_l)}\right\}
\sum_q L(q,0)\nonumber\\
&=&{N(0)DT\over (1+\Omega_n\tau)^2}\left\{-2\Omega_n^2\tau-
\Omega_n\left[\psi\left({1\over 2}+{|\Omega_n|\over 2\pi T}\right)
-\psi\left({1\over 2}\right)\right]+{(1+\Omega_n\tau)\Omega_n^2\over 4\pi T}
\left[\psi'\left({1\over 2}\right)
-\psi'\left({1\over 2}+{|\Omega_n|\over 2\pi T}\right)\right]\right\}
\sum_q L(q,0),
\end{eqnarray}
where we have explicitly carried out the $\varepsilon_l$ sum.
Upon analytically continuing $i\Omega_n\rightarrow\Omega$, dividing by $\Omega$, and 
taking the limit $\Omega\rightarrow 0$, the above expression gives zero result.
The net result is thus that the anomalous part of the DOS + MT
diagrams do not yield a divergent contribution.

Finally it only remains to show that there is no divergent contribution from the AL 
terms. Paradoxically, although this result does not appear to be in dispute, it is the
trickiest to prove. The method used is simple power-counting, applied to the analytical
continuation of the complete Matsubara sum. We need the complete sum because there is 
an anomalous region of Bose frequency, $\omega_m$, for
which the two superconducting propagators $L(q,i\omega_m+i\Omega_n)$ and 
$L(q,i\omega_m)$ have opposite signs of Matsubara frequency. We cannot therefore simply
take the static approximation where one or other superconducting propagator has zero
Matsubara frequency. Instead we must evaluate the two triangle blocks for general
$\omega_m$, and distinguish between the three summation regions: 
(i) $\omega_m+\Omega_n>0$, $\omega_m>0$; (ii) $\omega_m+\Omega_n>0$, $\omega_m<0$;
(iii) $\omega_m+\Omega_n<0$, $\omega_m<0$. Note that the two summation terms, $\omega_m=0$
and $\omega_m=-\Omega_n$, which possess one divergent fluctuation propagator, $L(q,0)$,
are both zero after analytic continuation $i\Omega_n\rightarrow\Omega+i0$, division by $\Omega$,
and taking the limit $\Omega\rightarrow 0$. It follows that when we analytically continue using
the contours shown in Fig. 2., we need not worry about contours passing through the poles. 

The contributions from regions (i) and (iii)
give identical results, and their sum is
\begin{equation}
\label{AL1def}
Q^{AL}_1(i\Omega_n)=-T\sum_{\omega_m>0}\sum_q {q^2\over d}B_1(i\omega_m,i\Omega_n)^2
L(q,i\omega_m)L(q,i\omega_m+i\Omega_n),
\end{equation}
where the $B_1(i\omega_m,i\Omega_n)$ are from the triangle blocks.
Upon replacing summation over $\omega_m$ by integration over $\omega$, and
analytically continuing $i\Omega_n\rightarrow\Omega+i0$, we get
\begin{equation}
\label{AL1cont}
Q^{AL}_1(\Omega)=-{1\over 4\pi i}\int_{-\infty}^{+\infty} d\omega\coth{(\omega/2T)}
\sum_q {q^2\over d}B_1(\omega,\Omega)^2 L(q,\omega)L(q,\omega+\Omega)
\end{equation}
For small $\omega$, $\Omega$, we can show that 
$B_1(\omega,\Omega)\approx\alpha\omega+\beta\Omega$, where $\alpha$ and $\beta$ are 
constants, so that for power-counting purposes Eq.~(\ref{AL1cont}) at $T=T_c$ becomes
(ignoring all irrelevant coefficients)
\label{AL1power}
\begin{equation}
Q^{AL}_1(\Omega)\sim
\int_{-\infty}^{+\infty} d\omega\coth{(\omega/2T)}\int d^dq q^2
{(\omega+\Omega)^2\over (q^2-i\omega)(q^2-i\omega-i\Omega)}
\end{equation}
The $O(\Omega)$ piece can be found by expanding either the numerator or denominator.
In both cases the behavior as $\omega\sim q^2\sim 0$ is $O(q^d)$, and hence there is
no infrared singularity for $d>0$.

The contribution from region (ii) has the form
\begin{equation}
\label{AL2def}
Q^{AL}_2(i\Omega_n)=-T\sum_{0>\omega_m>-\Omega_n}\sum_q 
{q^2\over d}B_2(i\omega_m,i\Omega_n)^2
L(q,i\omega_m)L(q,i\omega_m+i\Omega_n)
\end{equation}
which, upon replacing summation over $\omega_m$ by integration over $\omega$, gives
\begin{equation}
\label{AL2cont}
Q^{AL}_2(i\Omega_n)=-{1\over 4\pi i}
\left[\int_{-\infty}^{+\infty}d\omega-
\int_{-\infty-i\Omega_n}^{+\infty-i\Omega_n}d\omega\right]
\coth{(\omega/2T)}
\sum_q{q^2\over d}B_2(\omega,i\Omega_n)^2
L^A(q,\omega)L^R(q,\omega+i\Omega_n).
\end{equation}
Shifting variable in the second integral, $\omega\rightarrow\omega-i\Omega_n$,
analytically continuing $i\Omega_n\rightarrow\Omega+i0$, shifting the variable
back, $\omega\rightarrow\omega+\Omega$, dividing throughout by $\Omega$, and letting
$\Omega\rightarrow 0$ gives
\begin{equation}
\label{AL2final}
\lim_{\Omega\rightarrow 0}{Q^{AL}_2(\Omega+i0)\over\Omega}=
{1\over 8\pi iT}\int_{-\infty}^{+\infty}{d\omega\over\sinh^2{(\omega/2T)}}
B_2(\omega,0)^2\sum_q {q^2\over d} L^A(q,\omega)L^R(q,\omega)
\end{equation}
For small $\omega$ we can show that $B_2(\omega,0)=\gamma\omega$, where $\gamma$ is a
constant, so that for power-counting purposes Eq.~(\ref{AL2final}) at $T=T_c$ becomes
\begin{equation}
\label{A2power}
\lim_{\Omega\rightarrow 0}{Q^{AL}_2(\Omega+i0)\over\Omega}\sim
\int_{-\infty}^{\infty}{d\omega\over\sinh{(\omega/2T)}^2}
\int d^dq q^2 {\omega^2\over (q^2-i\omega)(q^2+i\omega)}
\end{equation}
The behavior as $\omega\sim q^2\sim 0$ is $O(q^d)$, and hence there is no infrared
singularity for $d>0$. We have therefore shown that there is no singular contribution
from the AL diagrams.

In conclusion we have shown that there are no superconducting fluctuation corrections to
the thermal conductivity above the transition temperature which are singular as $T$ approaches
$T_c$. The experimental features seen near $T_c$ must therefore have some other physical
explanation, such as reduced phonon scattering from normal state electrons. We hope that there will
be continued experimental interest in thermal conductivity near $T_c$ in one- and 
two-dimensional superconductors, of both the high-$T_c$ and low $T_c$ variety. In future work
we also intend to evaluate the non-singular fluctuation contributions to the thermal conductivity
to see if this can explain any of the experimental features (although, given their power-law
behavior, this seems unlikely).

\bigskip
\centerline {\bf ACKNOWLEDGEMENTS}
\medskip

We thank I.V. Lerner, G. Savona, A.A. Varlamov and I.V. Yurkevich for 
helpful discussions, and
S. Vishveshwara for bringing our attention to the question of fluctuation
contributions to thermal conductivity. We acknowledge support from 
the UK EPSRC.

\newpage

\begin{figure}[t]
\centerline{\psfig{figure=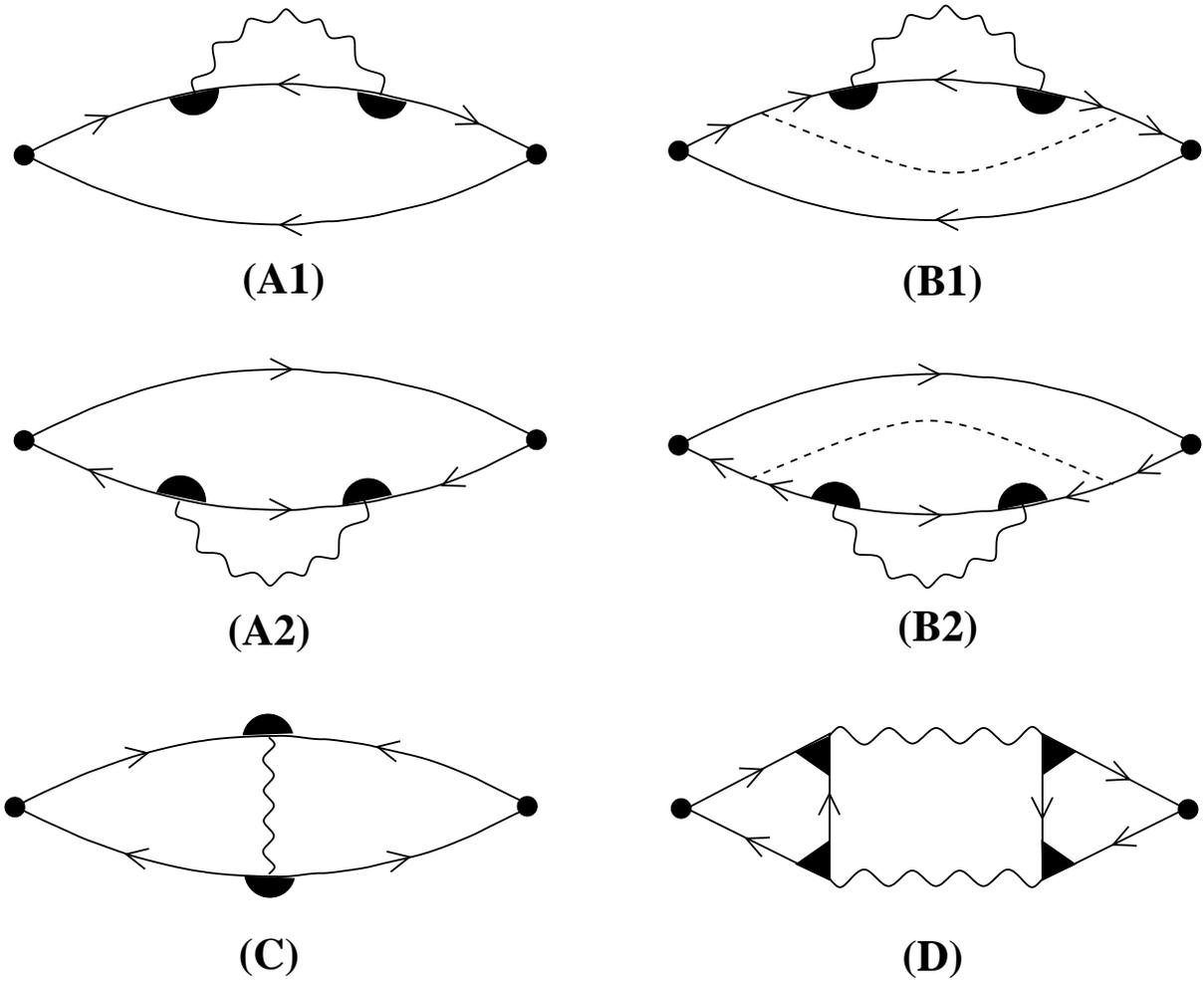,width=16cm}}
\medskip
\medskip
\medskip
\caption{
Feynman diagrams which give singular contributions to the heat-current response function.
Diagrams A and B are the density-of-states correction (DOS) diagrams; diagram C is the
Maki-Thompson (MT) diagram; diagram D is the Aslamazov-Larkin (AL) diagram.
}
\end{figure}

\newpage

\begin{figure}[t]
\centerline{\psfig{figure=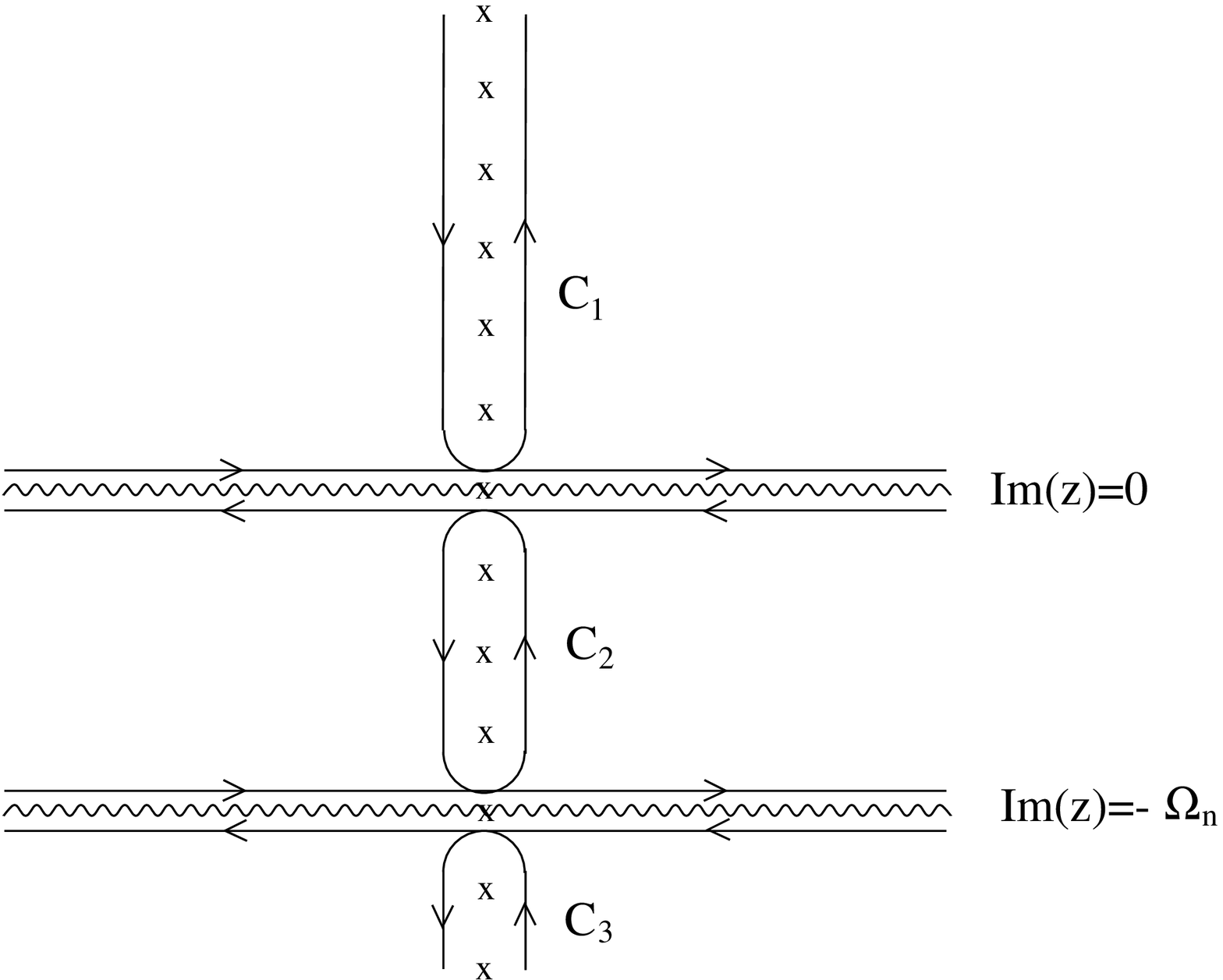,width=16cm}}
\medskip
\medskip
\medskip
\caption{
Contour required to perform sum over Matsubara frequencies $\omega_m$ in the AL diagram.
The branch cuts at $Im(\omega)=0$ and $Im(\omega)=-\Omega_n$ come from the fluctuation
propagators $L(q,i\omega_m)$ and $L(q,i\omega_m+i\Omega_m)$. The poles summed over fall into
three regions separated by the two branch cuts: (i) $\omega_m>0$; (ii) $0>\omega_m>-\Omega_n$;
(iii) $-\Omega_n>\omega_m$. These contours can be deformed to contours parallel to the real
axis as shown in the figure. Note that the poles which lie on the branch cuts yield no singular
contribution and can be ignored.
}
\end{figure}


\begin{references}

\bibitem{Lark01} A. Larkin and A. Varlamov, e-print cond-mat/0109177.

\bibitem{Skoc75} W.J. Skocpol and M. Tinkham, Rep. Prog. Phys. {\bf 38}, 
1049 (1975).

\bibitem{Cohn92} J.L. Cohn, E.F. Skelton, S.A. Wolf, J.Z. Liu and R.N. Shelton,
Phys. Rev. B {\bf 45}, 13144 (1992).

\bibitem{Hous96} M. Houssa, H. Bougrine, S. Stassen, R. Cloots, and M. Ausloos,
Phys. Rev. B {\bf 54}, R6885 (1996).

\bibitem{Hous97} M. Houssa, M. Ausloos, R. Cloots, and H. Bougrine, Phys. Rev.
B {\bf 56}, 802 (1997).

\bibitem{Sham00} G.A. Shams, J.W. Cochrane, and G.J. Russell, Physica C
{\bf 336} (2000).

\bibitem{Abra70} E. Abrahams, M. Redi and J.W.F. Woo, Phys. Rev. B {\bf 1},
208 (1970).

\bibitem{Varl90} A.A. Varlamov and D.V. Livanov, Zh. Eksp. Teor. Fiz. {\bf 98},
584 (1990) [Sov. Phys. JETP {\bf 71}, 325 (1990)].

\bibitem{Varl92} A.A. Varlamov, L. Reggiani, and D.V. Livanov, Phys. Lett. A
{\bf 165}, 369 (1992).

\bibitem{Vish01} S. Vishveshwara and M.P.A. Fisher, Phys. Rev. B {\bf 64},
134507 (2001).

\bibitem{Ussi02} I. Ussishkin, S.L. Sondhi, and D.A. Huse, e-print 
cond-mat/0204484.

\bibitem{Sav02} G. Savona, D.V.Livanov, R. Raimondi, and A.A. Varlamov,
e-print cond-mat/0207252.

\bibitem{Varl99} A readable discussion of the physical meaning of the various
fluctuation contributions to the electrical conductivity may be found in
A. Varlamov, G. Balestrino, E. Milani, and D. Livanov, Adv.
Phys. {\bf 48}, 655 (1999).

\bibitem{Wolf71} S. Wolf and B.S. Chandrasekhar, Phys. Rev. B {\bf 4}, 3014
(1971).


\end{references}
\end{document}